\documentclass[12pt]{iopart}
\usepackage[sc]{mathpazo}
\usepackage{helvet}
\usepackage[utf8]{inputenc}
\usepackage{graphicx}
\usepackage{floatrow}
\usepackage{xspace}
\usepackage{epstopdf}

\begin{document}

\title[GaN on Si by HVPE]
{Hydride Vapor Phase Epitaxy of GaN on Silicon Covered by Nanostructures}

\author{U Jahn$^{1}$, M Musolino$^{1}$, J Lähnemann$^{1}$, P Dogan$^{1}$,  S Fern\'{a}ndez Garrido$^{1}$, J F Wang$^{2}$, K Xu$^{2}$, D Cai$^{2}$, L F Bian$^{2}$, X J Gong$^{2}$ and H Yang$^{2}$}

\address{$^{1}$ Paul-Drude-Institut f\"{u}r
Festk\"{o}rperelektronik, Hausvogteiplatz 5-7, 10117 Berlin,
Germany}

\address{$^{2}$ Suzhou Institute of Nano-tech and Nano-bionics, Chinese Academy of Sciences, 398 Roushui Road, Suzhou City, 215123 P. R. China}

\begin{abstract}
Several ten~$\mu$m GaN have been deposited on a silicon substrate using a two-step hydride vapor phase epitaxy (HVPE) process. The substrates have been covered by AlN layers and GaN nanostructures grown by plasma-assisted molecular-beam epitaxy. During the first low-temperature (low-$T$) HVPE step, stacking faults (SF) form, which show distinct luminescence lines and stripe-like features in cathodoluminescence images of the cross-section of the layers. These cathodoluminescence features allow for an insight into the growth process. During a second high-temperature (high-$T$) step, the SFs disappear, and the luminescence of this part of the GaN layer is dominated by the donor-bound exciton. For templates consisting of both a thin AlN buffer and GaN nanostructures, a silicon incorporation into the GaN grown by HVPE is not observed. Moreover, the growth mode of the (high-$T$) HVPE step depends on the specific structure of the AlN/GaN template, where in a first case, the epitaxy is dominated by the formation of slowly growing facets, while in a second case, the epitaxy proceeds directly along the $c$-axis.

For templates without GaN nanostructures, cathodoluminescence spectra excited close to the Si/GaN interface show a broadening toward higher energies indicating a silicon incorporation on a high dopant level.
\end{abstract}


\footnote[0]{This is an author-created,un-copyedited version of an article accepted for publication in Semiconductor Science and Technology. IOP Publishing Ltd is not responsible for any errors or omissions in this version of the manuscript or any version derived from it. The Version of Record is available online at doi:10.1088/0268-1242/31/6/065018}

\maketitle

\section{Introduction}

In order to improve, e. g., the performance of nitride-based laser diodes, the density of threading dislocations penetrating into the corresponding layer systems need to be further reduced. This is a difficult task, when growing such device structures on foreign substrates, e. g., on sapphire, silicon carbide, or silicon due to a strain-related generation of extended defects.  Therefore, the interest in the fabrication of bulk GaN substrates as well as of free-standing GaN layers (FS-GaN) for a subsequent homo-epitaxial deposition of device structures is growing.
Hydride vapor phase epitaxy (HVPE) has already proven to be well suited for the fabrication of FS-GaN with threading dislocation densities below 10$^{4}$~cm$^{-2}$. \cite{Xu}
Commonly, sapphire wafers are used as substrates for the HVPE process to grow FS-GaN. \cite{Xu,Shibata,Lee1}

However, due to the low cost, available large wafer size, and well-established chemical etching techniques, silicon is very attractive to serve as a substrate material for the growth of FS-GaN.
So far, two key problems, which depend on each other, have prevented a successful use of silicon as a substrate for the HVPE  growth of FS-GaN. First, there is a high lattice mismatch (17$\%$) and a large difference in the thermal expansion coefficient (56$\%$) between GaN and silicon giving rise to cracking of the GaN during the growth as well as during the cooling of the GaN substrate from the growth temperature down to room temperature.
Second, silicon is not stable in the reactive ambiance and at the high temperatures applied during the HVPE process, where the reaction between silicon and the chlorine-based growth chemistry often results in a mixture of silicon, Ga, and N including SiN$_x$ phases (silicon-melt-back etching) instead of the intended GaN. 
\cite{Gu,Honda} 

AlN buffer layers and a two-step growth mode with first a low-temperature (low-$T$) step have been used to suppress the silicon-melt-back etching. Also selective-area growth (SAG), where the templates for the HVPE consist of a few 100~$\mu$m large growth windows to prevent cracking, has been applied. \cite{Nishimura}  However, the optimization of the protecting interlayer is very critical. For too thin AlN buffer layers, the reaction with silicon hinders an epitaxial growth of GaN. For too thick buffer layers, micro-cracks can appear during the growth, which again open a way for a reaction of Si with Ga or N. 
Thus, although the growth of a few ten~$\mu$m thick GaN layers on silicon by HVPE has already been reported about 10 years ago \cite{Lee2,Bessolov}, a long-term HVPE growth of GaN on silicon resulting in substrates with a thickness of several hundred~$\mu$m is still challenging.

Furthermore, the use of SAG or epitaxial lateral overgrowth techniques require micrometer masking, ex situ SiO$_2$ or SiN$_x$ deposition and multi-step photolithography, which increase the process complexity and costs.
Hersee $et~al. $\cite{Hersee} proposed a nano-heteroepitaxy technique, taking advantage of three-dimensional stress relief mechanisms, which is expected to result in a further improvement of GaN epilayers grown on foreign substrates. Based on this method, several authors have discussed a new non-lithographic nano-heteroepitaxy approach relying on self-assembled nanostructures in order to reduce the process steps and costs. \cite{Liang,Fu,Wang,Chan, Kwon}
Kwon $et~al.$ \cite{Kwon} have used a 50~nm thick sputtered AlN buffer on silicon, on which they have grown self-assembled nano-needles and -rods during a first low-$T$ HVPE step at 600 and 650~$^{\circ}$C, respectively. Subsequently, 0.5--2~$\mu$m thick GaN layers have been overgrown in a second HVPE step at 1040~$^{\circ}$C. The optical near-band gap emission of these GaN layers are red-shifted  with regard to the expected spectral position, the origin of which has not been analyzed in detail. Moreover, longer high-$T$ HVPE runs to obtain thick GaN layers have not been carried out.

In this work, we are reporting on the HVPE growth of thick GaN layers on self-assembled GaN nanostructures, the latter having been deposited on Si(111) substrates with AlN buffer layers using plasma-assisted molecular-beam epitaxy (MBE). The HVPE growth process is investigated by spatially and spectrally resolved cathodoluminescence (CL) spectroscopy in a scanning electron microscope (SEM) at cross-sections of the layers. By adding GaN nano-structures as nucleation sites and applying a two-step HVPE growth process, silicon incorporation into several ten~$\mu$m thick GaN layers could be prevented.

\section{Experimental details}

\subsection{Preparation of the templates}
Three templates have been prepared by MBE for the HVPE overgrowth to obtain several ten to several hundred~$\mu$m thick GaN layers on silicon. For sample S1, a commercial Kyma-Si/AlN template consisting of an n-type Si(111) wafer and a 200~nm thick Al-face AlN buffer layer have been used. On top of this template, GaN has been deposited by MBE at a temperature of 800~$^{\circ}$C setting the III/V flux ratio to 0.5. As a result, we obtained GaN nanostructures consisting of highly dense 0.6~$\mu$m high nano-walls and of taller nano-wires (NW) with a low density as shown in the SEM image of S1 in figure~\ref{fig1} (a) and indicated by white arrows in the top view and cross-sectional image of figure~\ref{fig1} (a). 
The very thin GaN NWs visible in figure~\ref{fig1} (a), which emerge from the GaN matrix, resemble a lot the ones observed by Fern\'{a}ndez Garrido $et~al.$ \cite{Garrido} on high quality Al-polar AlN layers grown on SiC (0001). For this reason we believe that the results obtained by Fern\'{a}ndez Garrido $et~al.$ are also valid for sample S1, that is the very thin GaN NWs are N-polar, whereas the GaN matrix is Ga-polar.     

The template of samples S2 and S4 consisted of an n-type Si(111) wafer and a 25~nm thick AlN buffer layer grown by MBE at a temperature of 680~$^{\circ}$C with a III/V flux ratio close to unity. This Si/AlN template was transferred to another MBE chamber to add GaN nanostructures, which have been grown at 780~$^{\circ}$C using a III/V flux ratio of 0.31. The resulting nanostructures consist of a rough parasitic layer, straight NWs with low density, and a large density of tilted nano-leafs [cf. SEM image of S2 and S4 in figure~\ref{fig1} (b)].

The template of the reference sample S3 consists of a Si(111) wafer and a 15~nm thick AlN layer grown by MBE at a temperature of 680~$^{\circ}$C setting the III/V flux ratio close to unity. As shown in the atomic force microscopy (AFM) image of figure~\ref{fig1} (c), the AlN layer of S3 is almost contineous with scattered pits. The layer is characterized by a dense distribution of small islands and a root-mean-square value of the roughness of about 0.8~nm. Further details about the growth parameters and microstructure of comparable AlN layers can be found elsewhere. \cite{Musolino}

\setcounter{figure}{0}
\begin{figure*}[h]
\floatbox[{\capbeside\thisfloatsetup{capbesideposition={right,center},capbesidewidth=10cm}}]{figure}[4cm]
{\includegraphics[width=6.4cm]{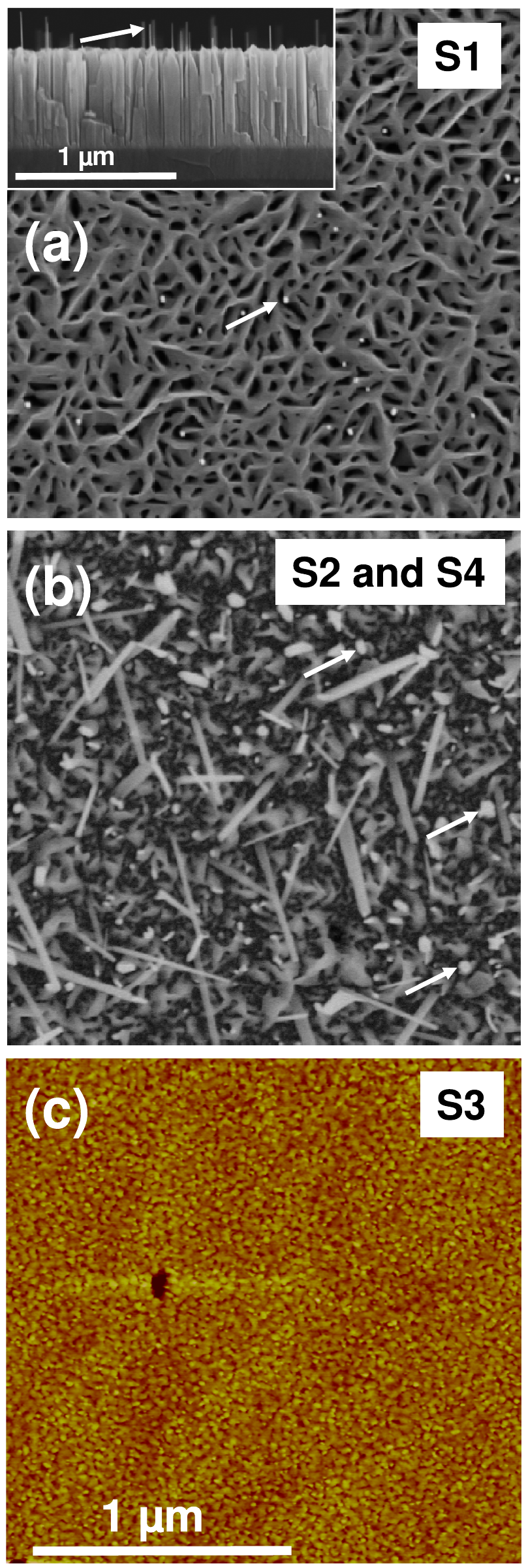}}
{\caption{(a) and (b): Top-view SEM images of the samples S1, S2, and S4 after the deposition of AlN buffer layers and GaN nanostructures on the Si(111) substrates by MBE. For S1, the inset shows a side view of the GaN nanostructure consisting of 600~nm high nanowalls and thin nanowires. For S1, S2, and S4, the arrows indicate the appearance of straight nanowires. (c): AFM map of the Si(111) substrate of S3 covered only by a 15~nm thick AlN layer.}
\label{fig1}}
\end{figure*}

\subsection{HVPE overgrowth}
The described templates showing very different surface structures were overgrown by thick GaN layers using a home-made horizontal reactor HVPE system. During the GaN growth, N$_{2}$ and H$_{2}$ were used as a carrier gas with a ratio of 1:1. The temperature of the Ga boat was set to about 850~$^{\circ}$C. The samples S1 to S3 were overgrown separately, but using the same conditions. We applied a two-step growth mode approach, which consisted of a low-$T$ (800~$^{\circ}$C) and a subsequent high-$T$ (950~$^{\circ}$C) step. During the first HVPE step, a lower temperature together with low HCl and NH$_{3}$ fluxes have been chosen in order to prevent the evolution of silicon-melt-back etching. The HCl and NH$_{3}$ fluxes amounted to 0.01 and 0.6~slm during the first step and to 0.04 and 1.8~slm during the second step. For S1 and S2--S4, the growth time of the low-$T$ step amounted to 20 and 50 minutes, respectively. The high-$T$ step lasted for 30 minutes for all samples. In the case of sample S4, an additional 360~minutes long high-$T$ step has been performed at 1063~$^{\circ}$C in order to obtain a GaN substrate of several hundred~$\mu$m thickness.
In order to provide a more clear overview of the layer parameters and growth conditions, we summarize the thickness values of the used AlN buffers and of the HVPE GaN layers as well as the HVPE growth temperatures and corresponding growth times of the samples S1 to S4 in Table~\ref{tab1}. 

\begin{table}[ht]
\caption{Thickness of the AlN buffer layer ($d_{AlN})$ grown by MBE, growth temperature and time of the low-$T$  HVPE step ($T_{G1}$ and $t_{G1}$), growth temperature and time of the high-$T$  HVPE steps ($T_{G2}$ and $t_{G2}$, $T_{G3}$ and $t_{G3}$), thickness of the GaN layer after the low-$T$ HVPE step ($d_1$), and total thickness of the HVPE GaN ($d_{total}$) for the samples S1 to S4.}
\label{tab1}
\begin{center}
\begin{tabular}{|l|cccc|}
\hline
Sample & S1 & S2 & S3 & S4 \\
\hline
$d_{AlN}$~(nm) & 200 & 25 & 15 & 25 \\ 
\hline
$T_{G1}$~($^{\circ}$C), $t_{G1}$~(min.) & 800, 20 & 800, 50 & 800, 50 & 800, 50 \\ 
$d_1$~($\mu$m) & 5 & 10 & --- & --- \\ 
\hline
$T_{G2}$~($^{\circ}$C), $t_{G2}$~(min.) & 950, 30 & 950, 30 & 950, 30 & 950, 30 \\ 
$T_{G3}$~($^{\circ}$C), $t_{G3}$~(min.) & --- & --- & --- & 1063, 360 \\
\hline 
$d_{total}$~($\mu$m) & 25 & 40 & 25 & 600 \\ 
\hline
\end{tabular}
\end{center}
\end{table}

\subsection{Analysis}
The morphology of all of these samples was imaged by means of a field-emission SEM or by means of an AFM. The spatial distribution of the luminescence intensity as well as luminescence spectra from the cross-sections of the overgrown GaN layers were measured using a CL system attached to the SEM. The cross-sections were prepared by using an Ar-ion cutting machine.
SEM and monochromatic CL images were acquired simultaneously for an accurate assignment of the local origin of the CL. A He-cooling stage allowed to vary the sample temperatures between 6 and 300 K. The CL system was operated with a photomultiplier tube for both monochromatic imaging and to record CL spectra. For the acquisition of both CL spectra and CL images, the acceleration voltage and the current of the electron beam were chosen to be 5~kV and about 1~nA, respectively.

\section{Results and discussion}

\subsection{Surface morphology and luminescence spectra of S1--S3}
After the HVPE growth, the wafers were slowly cooled down from the growth temperature to room temperature over a period of 4 hours. Nevertheless, cracks appeared within the GaN layers of S1 to S3 caused by the large tensile strain due to the large difference of the thermal expansion coefficients between silicon and GaN. The typical crack density amounts to several cracks per cm$^{2}$. For S1 to S3, the thickness of the HVPE GaN layer amounts to 25 to 35~$\mu$m. As confirmed by an energy dispersive x-ray (EDX) analysis at  the cross-sections of these HVPE GaN layers (not shown), silicon-melt-back etching was successfully suppressed during the HVPE process. Figure \ref{fig2} shows the SEM images of the surface of the HVPE GaN layers. From sample to sample, the SEM images exhibit a varying density and size of V-pits. The V-pit density amounts to 6.5$\times$10$^6$~cm$^{-2}$ for S1, 3$\times$10$^6$~cm$^{-2}$ for S2 and 1.4~$\times$10$^7$~cm$^{-2}$ for S3. The origin of these V-pits is discussed below. Using photoelectron diffraction to determine the polarity of the HVPE-GaN layers \cite{Romanyuk}, it has been confirmed that the layers are Ga-polar (not shown here).

\setcounter{figure}{1}
\begin{figure*}[h]
\floatbox[{\capbeside\thisfloatsetup{capbesideposition={right,center},capbesidewidth=10cm}}]{figure}[4cm]
{\includegraphics[width=6.4cm]{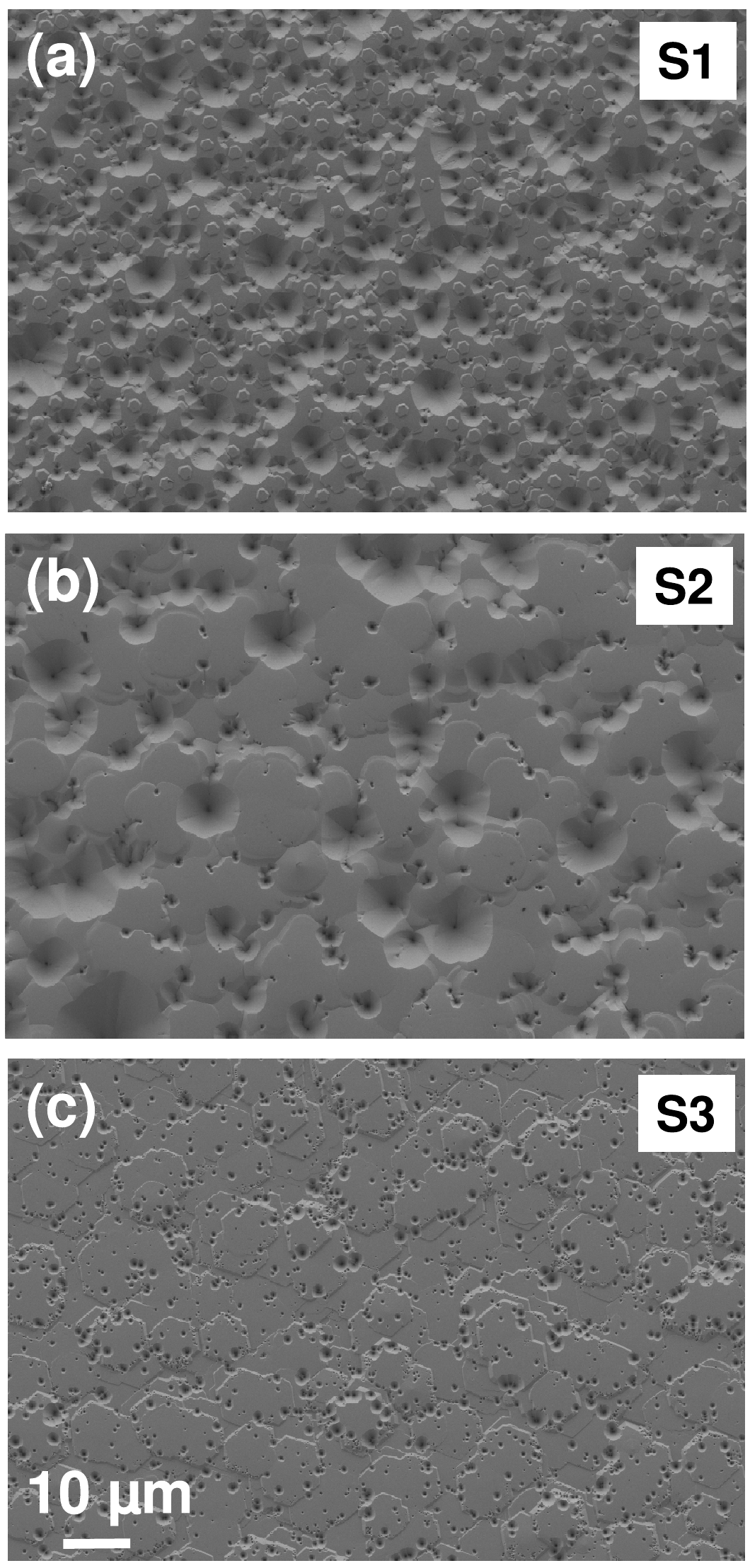}}
{\caption{(a)--(c): Top-view SEM images of the samples S1--S3, respectively, after the overgrowth of the templates shown in Fig.~\ref{fig1} by several ten~$\mu$m of GaN using HVPE.}
\label{fig2}}
\end{figure*}

In order to investigate the spatial distribution of the luminescence intensity within the HVPE layers, in particular, along the growth direction, we performed CL measurements at the cross-sections prepared by ion cutting.
Figures \ref{fig3}(a) and (b) show normalized CL spectra at 6~K, which have been excited within regions located at the upper part of the cross-sections and within regions close to the AlN interlayer of S1--S3, respectively, as indicated by the solid rectangles in the inset of figure\ref{fig3}(a).

\setcounter{figure}{2}
\begin{figure}[h]
\includegraphics*[width=10cm]{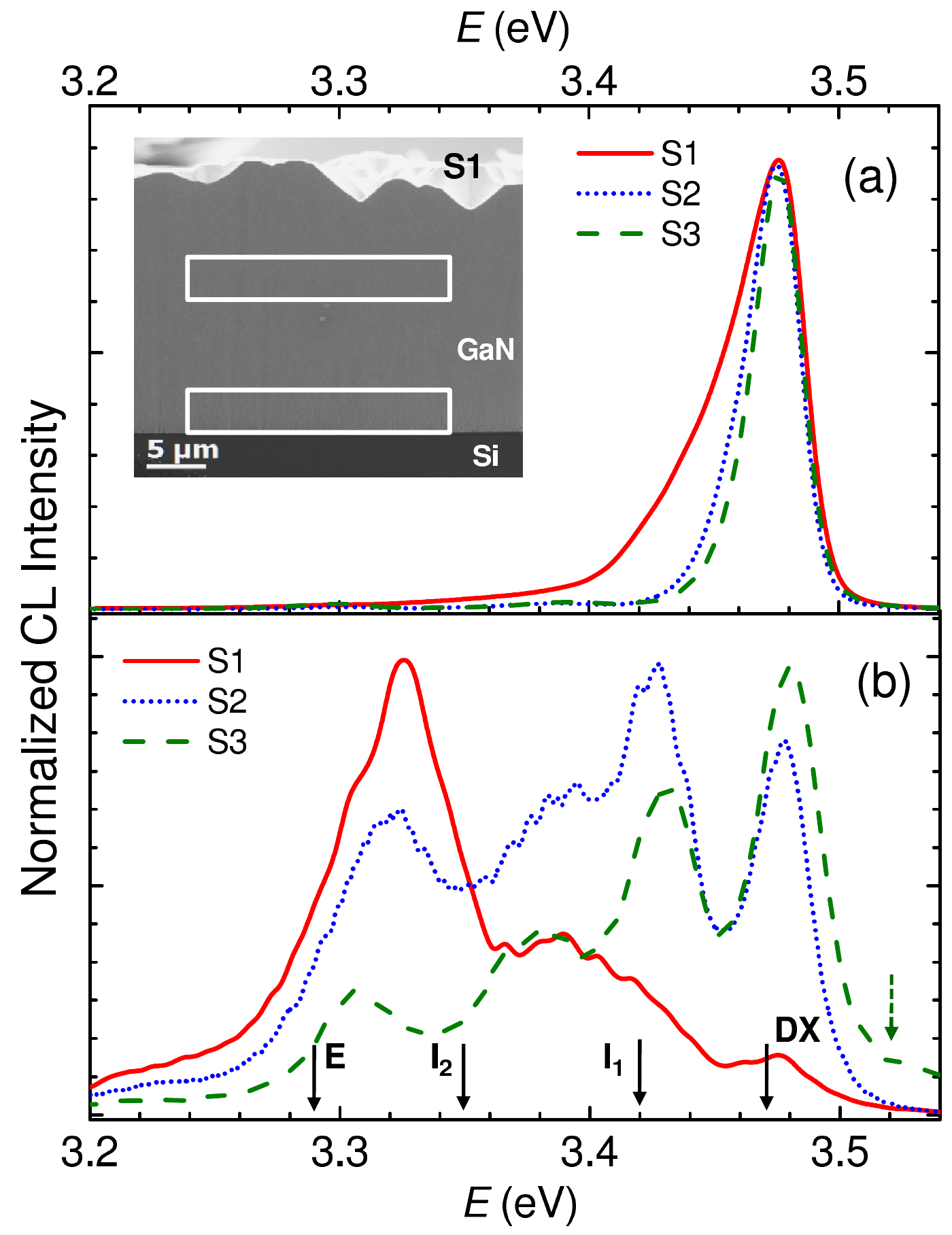}
\caption{CL spectra of the samples S1, S2, and S3 at 6~K, which have been excited within the upper (a) and lower (b) part of the corresponding cross-sections as marked by the solid rectangles in the SEM image of the inset in (a). In (b) E, I$_2$, I$_1$, and DX denote the spectral positions of the excitons bound to extrinsic, I$_2$- as well as I$_1$-type basal plane stacking faults, and the donor-bound exciton in GaN, respectively.}
\label{fig3}
\end{figure}

Luminescence spectra indicate sensitively the presence of extended defects such as inversion domain boundaries (IDB) and stacking faults. According to the recent findings of Auzelle $et~al.$ \cite{Auzelle}, the presence of IDBs in the HVPE-GaN layers should result in relatively intense luminescence at a spectral position of about 3.45 eV. However, such spectral line does not seem to be present in the CL spectra of samples S1--S3, as shown in figure \ref{fig3}, or at least its intensity is very weak. The absence of the CL line at 3.45 eV indicates that the HVPE-GaN exhibits a uniform Ga-polarity. 

Furthermore, the CL spectra of the upper part of all HVPE GaN layers are dominated by the donor-bound exciton (DX) recombination centered at about 3.475~eV. However, the spectrum of S1 is significantly broadened toward the low-energy side. For the lower part --- close to the AlN/GaN interface --- the CL spectra of S1 and S2 are dominated by luminescence lines centered at energy values below 3.44~eV. We assign these lines to excitons bound to basal plane stacking faults (I$_1$ SF, I$_2$ SF, and E SF).  \cite{Lahnemann1, Lahnemann2} For sample S3, the CL spectrum of the lower part of the cross-section shows also significant contributions of SF-related luminescence, which are not as dominant as in S1 and S2. In contrast to S1 and S2, the CL spectrum of the lower part of S3 clearly shows a shoulder at an energy value exceeding the one of the GaN band edge [cf. dashed arrow in figure~\ref{fig3}(b)]. As will be discussed below, the origin of this high-energy shoulder could be a high unintentional doping of the GaN layer close to the Si/GaN interface due to silicon incorporation.

The spatial origin of the various parts of the CL spectra can be precisely determined by acquiring CL maps for the corresponding CL detection energies $E_{d}$. The ion-milled edges of the samples show smooth surfaces [cf. inset of figure~\ref{fig3}(a)] allowing for a mapping of the CL intensity without artifacts caused by a rough topography due to a typically non-uniform cleavage of GaN. 

\subsection{Spatial CL distribution at the cross-section of S1}
Figure~\ref{fig4}(a) shows  the CL map of the cross-section of sample S1 recorded for $E_{d}=3.475$~eV, which reveals that the donor-bound exciton emission originates from an about 20~$\mu$m thick upper region of the HVPE GaN, while the about 5~$\mu$m thick bottom layer does not show any significant donor-bound exciton CL. However, this region at the bottom of the HVPE GaN shows significant emission for values of $E_{d}$ lower than 3.45~eV as shown in figures~\ref{fig4}(b)--4(d), which we assign to SF-related luminescence. Thus, the CL maps clearly indicate two sharply separated regions along the growth direction, where the SF-related luminescence is found only within a 5~$\mu$m thick bottom layer, while near-band gap luminescence originates mainly from the upper, thick part of the HVPE GaN layer. The near-band gap luminescence of the upper region shows an inhomogeneous distribution with dark vertical lines and bands resulting from non-radiative recombination at threading dislocations and/or grain boundaries, the density of which decreases along the growth direction.

\setcounter{figure}{3}
\begin{figure}[h]
\includegraphics*[width=10cm]{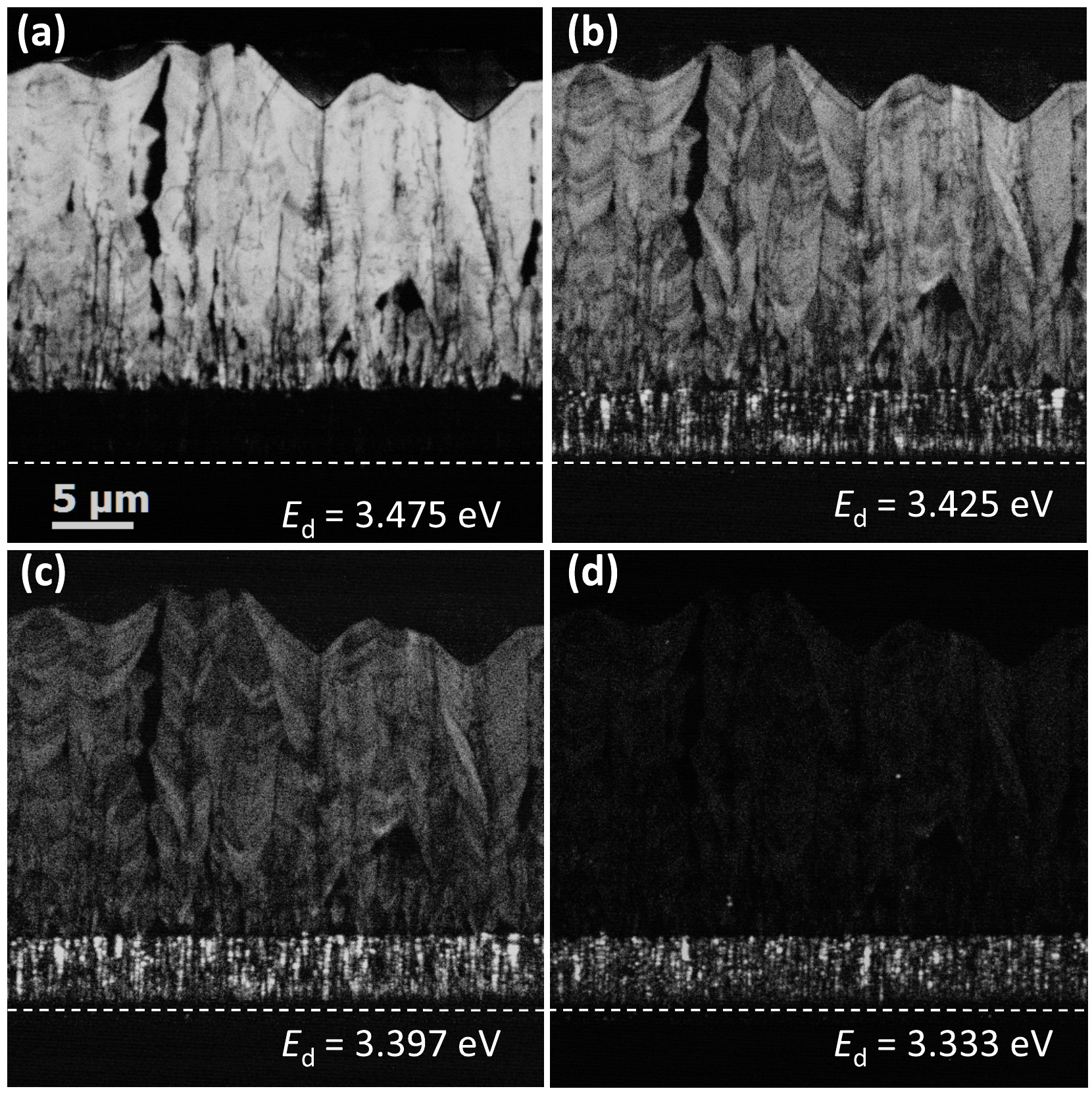}
\caption{CL images, all recorded on the same region of the cross-section of sample S1, for different CL detection energies $E_{d}$. $E_{d}=3.475$~eV corresponds to the donor-bound exciton in GaN. The dashed lines mark the GaN/Si interface.}
\label{fig4}
\end{figure}

Since the formation of SFs in GaN is known to be promoted at low growth temperatures \cite{Renard} as well as for growth along directions different from the $c$-direction \cite{Mei,Corfdir}, we conclude that the 5~$\mu$m thick SF-rich bottom region of the GaN layer is a result of the epitaxy during the first, low-$T$ HVPE step. Due to the typically high growth rate of the HVPE process, a temperature of 800~$^{\circ}$C is not sufficiently high to prevent instabilities of the crystallographic phase leading to the evolution of SFs or even of inclusions of the cubic phase of GaN. Moreover, at the beginning of the low-$T$ growth step, the nucleation of the HVPE GaN on the nanostructures is followed by lateral growth, until coalescence is completed,  with the latter also being known to lead to the formation of SFs. \cite{Lahnemann1}

Note that both the CL spectrum of figure~\ref{fig3}(a) and the CL map of figure~\ref{fig4}(b) of sample S1 show significant emission for $E_{d}=3.425$~eV also within the upper thick part of the HVPE GaN. However, the spatial distribution of the emission at this energy differs significantly from the one of the region at the bottom of the HVPE GaN, and, in contrast to the SF emission, comes from an area also showing DX emission, thus indicating a different origin of this emission. We clearly observe large V-shaped and small column-like contrasts of the CL mapping throughout the upper and lower regions of the layer, respectively. Figure~\ref{fig5} displays a magnified CL image of the lower part of the cross-section of sample S1 for a detection energy $E_{d}=3.425$~eV. At the bottom, the low-$T$ HVPE GaN shows CL features, which are arranged in columns and are elongated along the basal plane. The latter observation supports the assignment of this emission to SF-related luminescence. Directly from the interface to the low-$T$ HVPE GaN, the CL map of the high-$T$ HVPE GaN is characterized by V-shaped features as indicated by dashed lines. Specifically, the apex of these features originates at the boundary between two of the columns situated beneath. As described recently by Lee $et~al.$, \cite{Lee3,Lee4} V-shaped intensity distributions in CL maps of the cross-section of thick HVPE GaN layers are caused by the evolution of $\{10\bar{1}1\}$ facets initiated by threading dislocations. These facets, which form the observed V-pits, are nitrogen-terminated and promote the incorporation of oxygen atoms during growth. \cite{Lee4, Bu, Wei1} Since oxygen atoms give rise to a donor level with a donor-to-valence band transition at about 3.42~eV \cite{Chung,Chen}, the near-band edge CL line of the upper region of S1 is broadened toward the low-energy side, and the motion of the growth front of the $\{10\bar{1}1\}$ facets can be clearly imaged for $E_{d}=3.425$~eV. According to  Lee $et~al.,$ \cite{Lee3} the V-pits can get filled during the growth by the evolution of faster growing facets at the bottom of the $\{10\bar{1}1\}$-facetted V-pits giving rise to the observation of V- and U-shaped forms with varying opening angles. In figure~\ref{fig5}, the marked angles 1 and 2  amount to 62 and 55$^{\circ}$ with respect to the basal plane, respectively.

\setcounter{figure}{4}
\begin{figure}[h]
\includegraphics*[width=7cm]{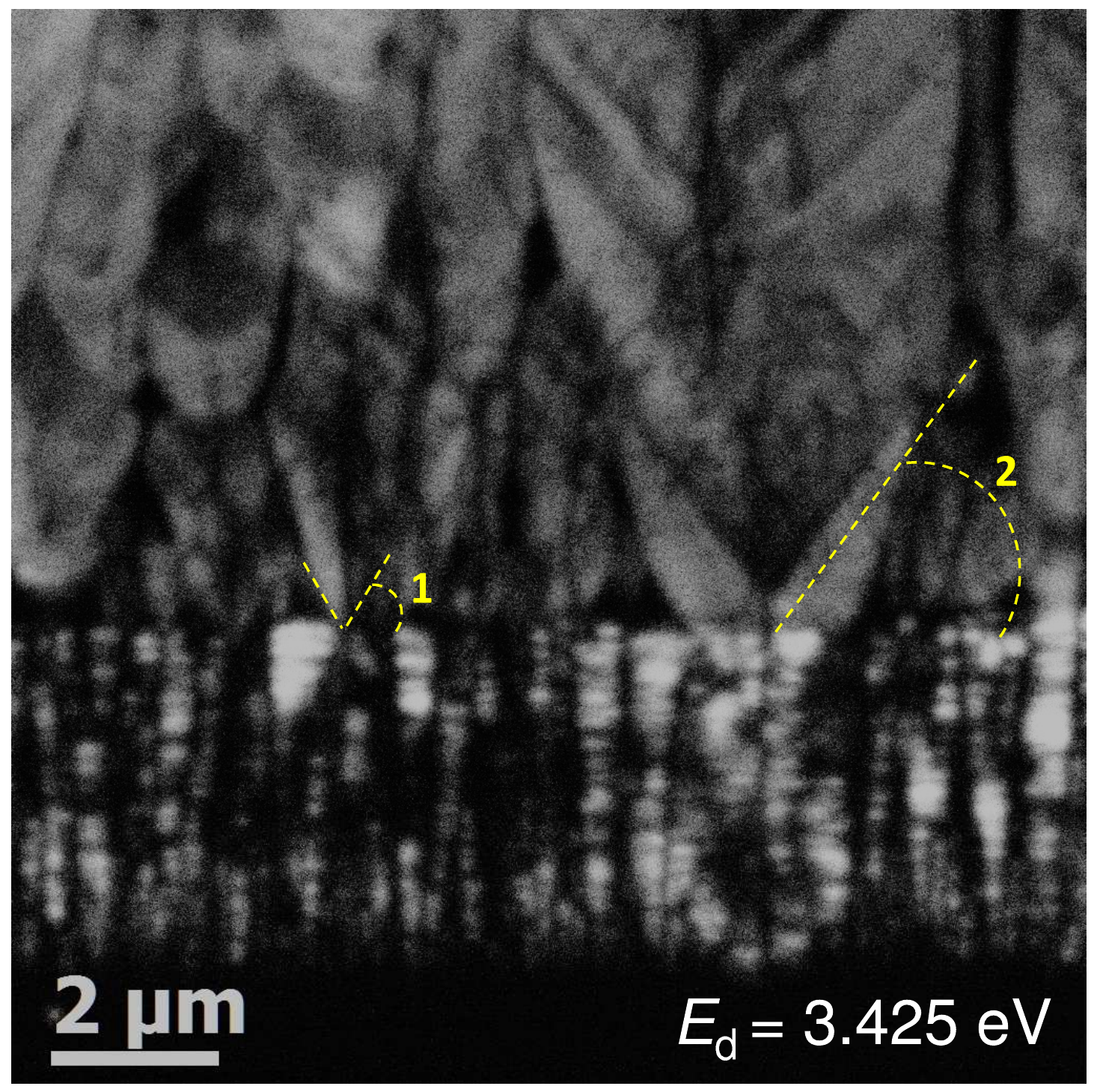}
\caption{Magnified CL image of the lower part of the cross-section of sample S1 for a detection energy $E_{d}=3.425$~eV. At the bottom, the firstly grown GaN shows SF-related CL features, which are  arranged as columns. Just above this layer, the CL map is characterized by V-shaped features as indicated by the dashed lines. The marked angles 1 and 2 amount to 62 and 55$^{\circ}$, respectively.}
\label{fig5}
\end{figure}

For the HVPE overgrowth of S1, we summarize the experimental results as follows. During the low-$T$ HVPE step, a columnar growth occurs with a mean column diameter of about 0.3~$\mu$m. At the very beginning of the subsequent high-$T$ HVPE step, V-pits evolve, the growth dynamics of which are governed by facet growth leading to an increase of the concentration of incorporated impurity atoms such as oxygen.  

\subsection{Spatial CL distribution at the cross-section of S2}
Figures~\ref{fig6}(a) and 6(b) show CL images of identical regions from the cross-section of sample S2 for CL detection energies of $E_{d}=3.475$ and 3.425~eV, respectively. Again, we observe two distinct parts of the HVPE GaN representing the  HVPE grown at low $T$ and high $T$, where the optical emission of the former is related to basal plane SFs, while the emission of the latter is related to the donor-bound exciton in GaN. Moreover, the CL distribution mapped at 3.425~eV is very weak throughout the whole upper part of the HVPE GaN and does not show any significant features such as V-shaped patterns.  An enlarged section of the superposition of the CL maps of figures~\ref{fig6}(a) and 6(b) is shown  in figure~\ref{fig6}(c) for a direct correlation. In the lower part of figure~\ref{fig6}(c), the CL intensity forms stripe segments oriented along the basal plane confirming again their basal plane SF-related origin. These stripe segments are arranged within columns with an average diameter of 2~$\mu$m, the borders of which are continued within the upper part of figure~\ref{fig6}(c) (cf. dashed arrows) indicating a columnar growth of the GaN along the $c$-direction during the high-$T$ HVPE step. Some of the dark border lines are disappearing along the growth direction due to coalescence of neighboring columns. We do not observe any indication of the evolution of  $\{10\bar{1}1\}$ or other slowly growing facets, i. e., of a V-pit-like growth at this stage of the epitaxy. Nevertheless, V-pits are also present at the surface of S2 as visible in figure~\ref{fig6}(a) and shown in figure~\ref{fig2}(b). These V-pits were, however, formed within a region close to the surface, where either a bundle of or single dark-line defects are found at their apex as shown in the CL image of figure~\ref{fig7}. The indicated angle of 62$^{\circ}$ corresponds to  $\{10\bar{1}1\}$ facets.

\setcounter{figure}{5}
\begin{figure}[h]
\includegraphics*[width=10cm]{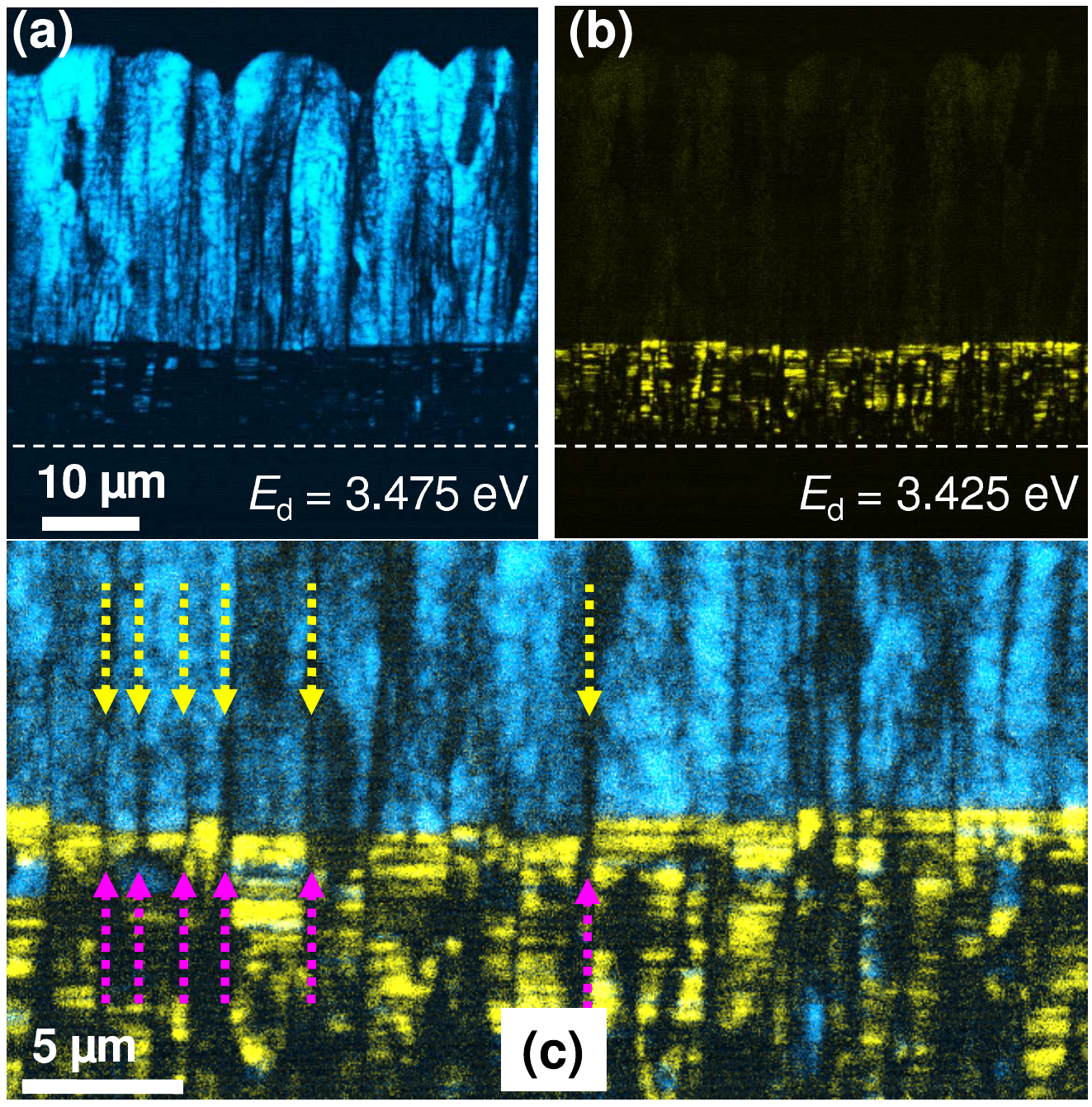}
\caption{CL images taken on identical regions of the cross-section of sample S2 for CL detection energies of (a) $E_{d}=3.475$ and (b) 3.425~eV,  which correspond to the donor-bound exciton and exciton bound to the I$_1$-type SF in GaN, respectively. The dashed lines mark the GaN/Si interface. (c) magnified section of a superposition of (a) and (b). In (c), the dashed upwards showing arrows mark the boundaries of the column-arranged SF-features at the bottom of the GaN layer, which continue into the upper part of the GaN (downwards showing arrows), thus indicating a columnar growth along the $c$-direction.}
\label{fig6}
\end{figure}

A reason for the appearance of the V-pits at the surface could be a non-uniform decomposition of the GaN during cooling down to room temperature after the HVPE has been terminated. Since the cooling rate is usually low in order to minimize cracking of the layer, the sample remains for a long time at high temperatures. If the decomposition rate is higher at positions, where threading dislocations penetrate the surface, compared with the surrounding surface regions, V-pits can be formed at dislocation positions as shown in figure~\ref{fig7}.

\setcounter{figure}{6}
\begin{figure}[h]
\includegraphics*[width=7cm]{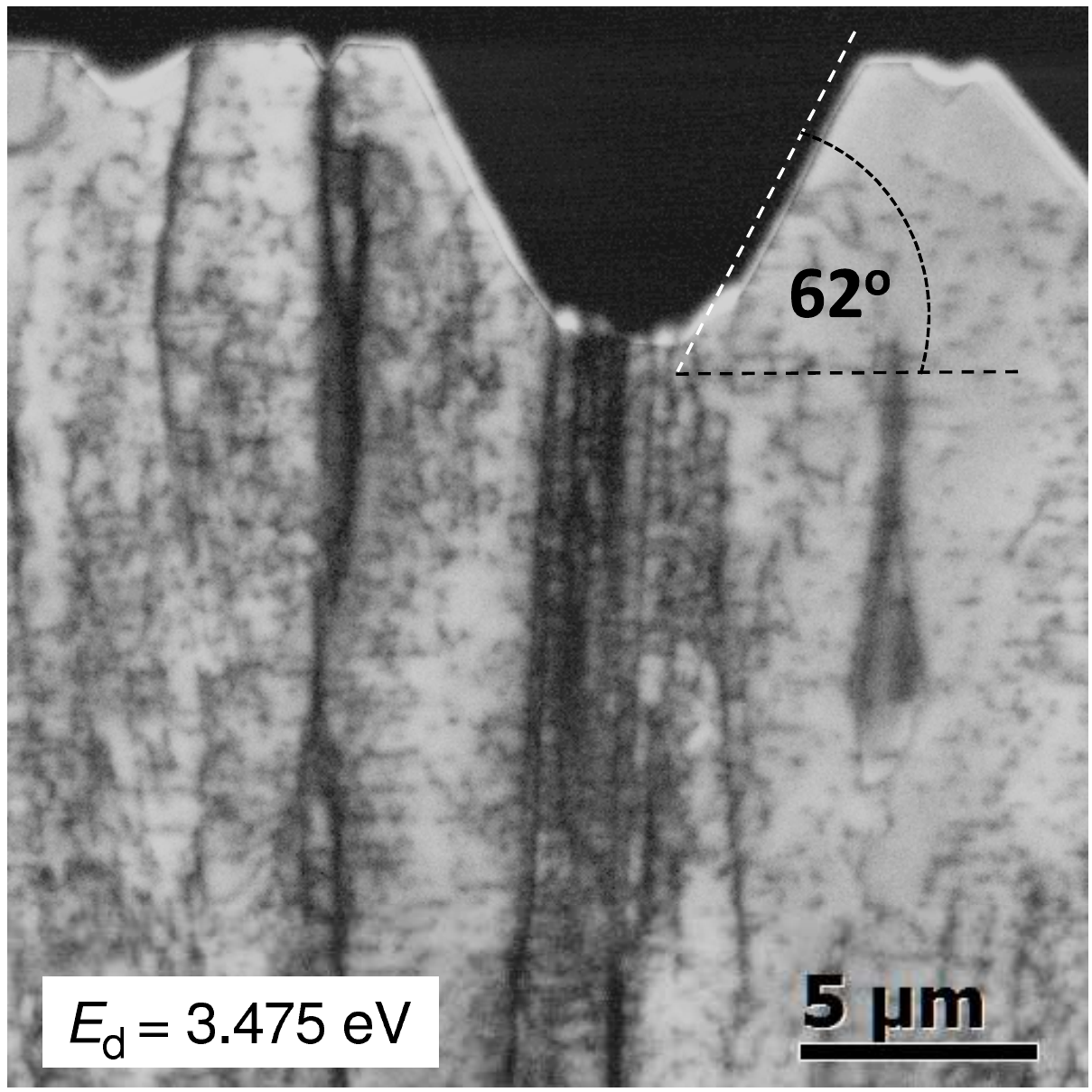}
\caption{CL image of the topmost part of the cross-section of sample S2 for a detection energy of $E_{d}=3.475$~eV showing V-pits at the surface. The indicated angle corresponds to the $\{10\bar{1}1\}$ facet of wurtzite GaN.}
\label{fig7}
\end{figure}

Comparing the above discussed CL data of S1 and S2, we assume that the different morphology of the underlying MBE-grown nanostructure (cf. figure~\ref{fig1}), where the density of the nanowalls of S1 is much higher than the density of NWs or nanoleafs of S2, resulted in a columnar growth with very different mean column diameters (0.3~$\mu$m for S1 and 2~$\mu$m for S2) during the low-$T$ HVPE step.  According to Kwon $et$ $al.$ \cite{Kwon}, a larger distance of the nanostructures serving as the template for a HVPE overgrowth promotes a lateral epitaxy and therefore results in larger grains during the first HVPE step. The very different mean column diameters formed during the low-$T$ HVPE step resulted in a different growth dynamics for S1 and S2 during the subsequent high-$T$ HVPE step. While in the case of S1, a  $\{10\bar{1}1\}$ faceted growth dominates, we observe a columnar growth along the $c$-direction throughout the whole GaN layer for S2.

\subsection{Spatial CL distribution at the cross-section of S3}
For the reference sample S3, the CL distribution of the cross-section differs significantly from the ones of S1 and S2. In figure~\ref{fig8}(a), the cross-section CL image of S3 obtained for $E_{d}=3.475$~eV shows near-band gap optical emission throughout the whole thickness without a clear distinction between the low-$T$ and high-$T$ HVPE growth regions. A few dark horizontal lines appear close to the bottom of the GaN layer. Figure~\ref{fig8}(b) depicts low-$T$ CL spectra measured within the upper part of the cross-section of S3 and directly at the AlN/GaN interface. The latter is significantly broadened and blue-shifted compared with the spectrum obtained far away from the interface. Correspondingly, the CL maps of  figures.~\ref{fig8}(c) to 8(e) show bright luminescence features within an 1 to 2~$\mu$m thick interface region for the whole considered spectral range even for  $E_{d}=3.556$~eV, which clearly exceeds the band edge energy of GaN.  Above this interface region of the GaN layer, we observe some bright, SF-related stripe segments in the CL map for $E_{d}=3.428$~eV of figure~\ref{fig8}(e). The broadening and blue-shift of the CL spectrum obtained close to the interface indicate a strong doping of the near-interface region most probably by silicon atoms from the substrate.

\setcounter{figure}{7}
\begin{figure}[h]
\includegraphics*[width=12cm]{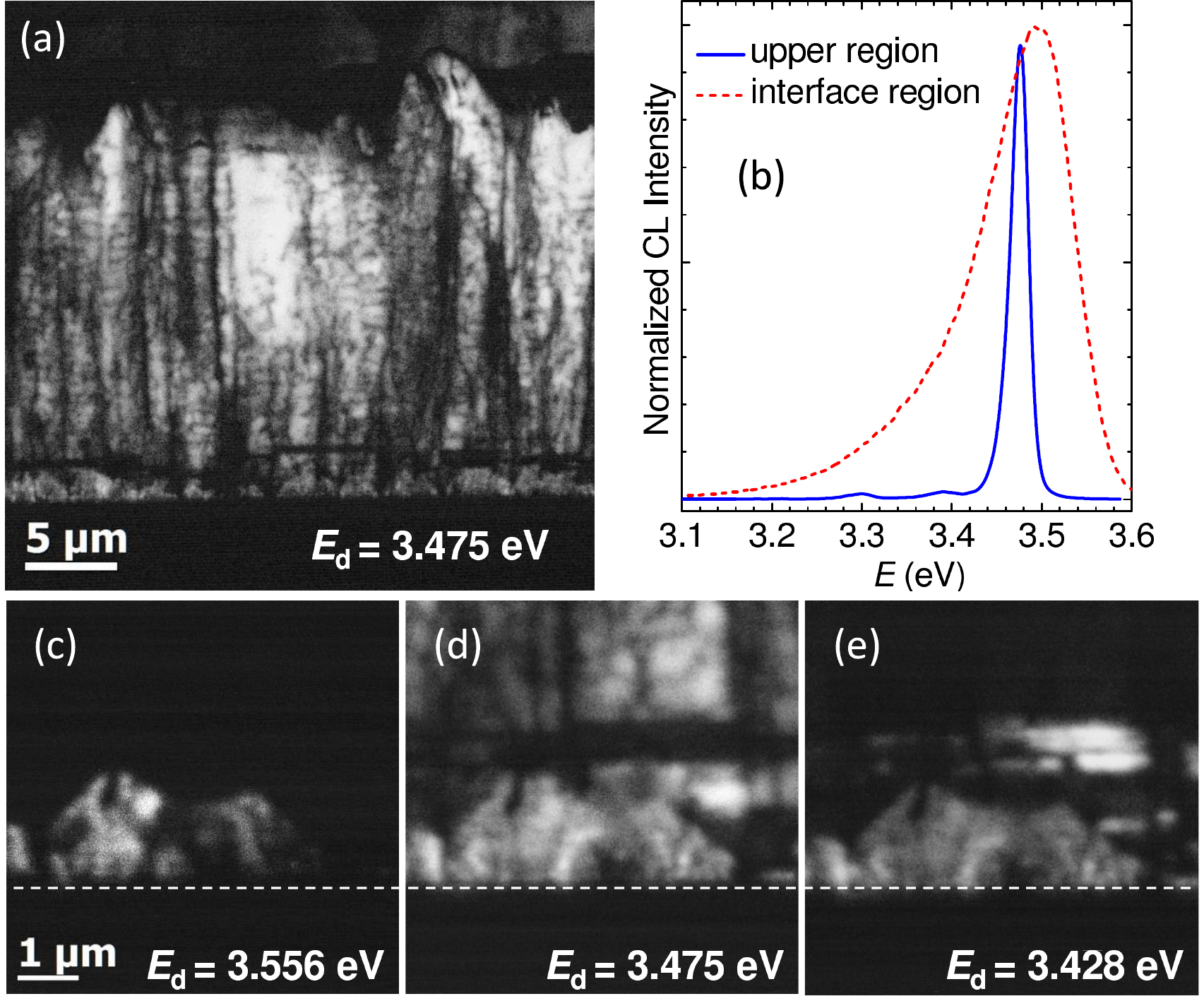}
\caption{(a) CL image of the cross-section of S3 for a CL detection energy of $E_{d}=3.475$~eV. (b) Low-$T$ CL spectra excited at the upper part of the cross-section (solid line) and at the AlN/GaN interface (dashed line) of S3. (c)--(e): Magnified CL images of a cross-section region close to the AlN/GaN interface of sample S3 differing in the CL detection energy as indicated. The energy value $E_{d}=3.556$~eV exceeds the band edge energy of GaN.  $E_{d}=3.475$~eV and  $E_{d}=3.428$~eV correspond to the donor-bound exciton and exciton bound to the I$_{1}$-type stacking fault, respectively. The dashed line marks the interface to the silicon substrate.}
\label{fig8}
\end{figure}

Since the AlN buffer layer exhibits a certain roughness, its real thickness can be locally smaller than the nominally expected 15~nm. Moreover, it is known from transmission electron microscope investigations that our MBE grown AlN buffers are closed, but not single crystalline layers, and therefore exhibit grain boundaries. The latter could serve as fast diffusion paths for Si \cite{Jakiela} leading to a local incorporation of silicon into the GaN layer on a strong dopant level. A silicon concentration on a dopant level cannot yet be detected by EDX and is probably not a result of a silicon-melt-back etching, which usually leads to a much higher silicon content. We suppose that due to the HVPE growth directly on top of the uncovered AlN buffer layer, a locally high silicon doping of the GaN takes place leading to a broad and blue-shifted optical emission spectrum, which we have also observed for AlN buffer layers as thick as 200~nm for a 6 hours lasting HVPE process at 800~$^{\circ}$C. The appearance of SF-related CL features just above this interface region could also be a signature of a strong silicon doping. \cite{Molina}

\subsection{Comparative discussion (S1--S3)}
In figure~\ref{fig9}, we compare the CL spectra excited at the upper part of the cross-sections of S1 to S3 at room temperature. S1 shows the highest CL intensity, but also the largest linewidth (FWHM$=130$~meV). Both the higher intensity and the larger FWHM are probably a result of doping due to a growth controlled by $\{10\bar{1}1\}$ or other slowly growing facets. The maximum CL intensity of S2 is about 10$\%$ lower than the one of S1, whereas the linewidth of the spectrum is significantly smaller (FWHM$=85$~meV) and is close to the linewidth of a corresponding spectrum obtained from a FS-GaN substrate, which exhibits a dislocation density as low as $10^6$~cm$^{-2}$. The linewidth of the CL spectrum of S3 is equal to the one of S2, but the CL intensity is about 60$\%$ lower than that of S2, which is probably due to a higher density of dark-line defects in S3 as compared with S2. This difference in the dark-line density can hardly be quantified by means of the CL maps shown in  figures.~\ref{fig6} and \ref{fig8}, but can be indirectly concluded by comparing the V-pit density at the surface, which amounts to $3\times10^6$~cm$^{-2}$ for S2 and $1.4~\times10^7$~cm$^{-2}$ for S3 (cf. figure~\ref{fig2}).

\setcounter{figure}{8}
\begin{figure}[h]
\includegraphics*[width=10cm]{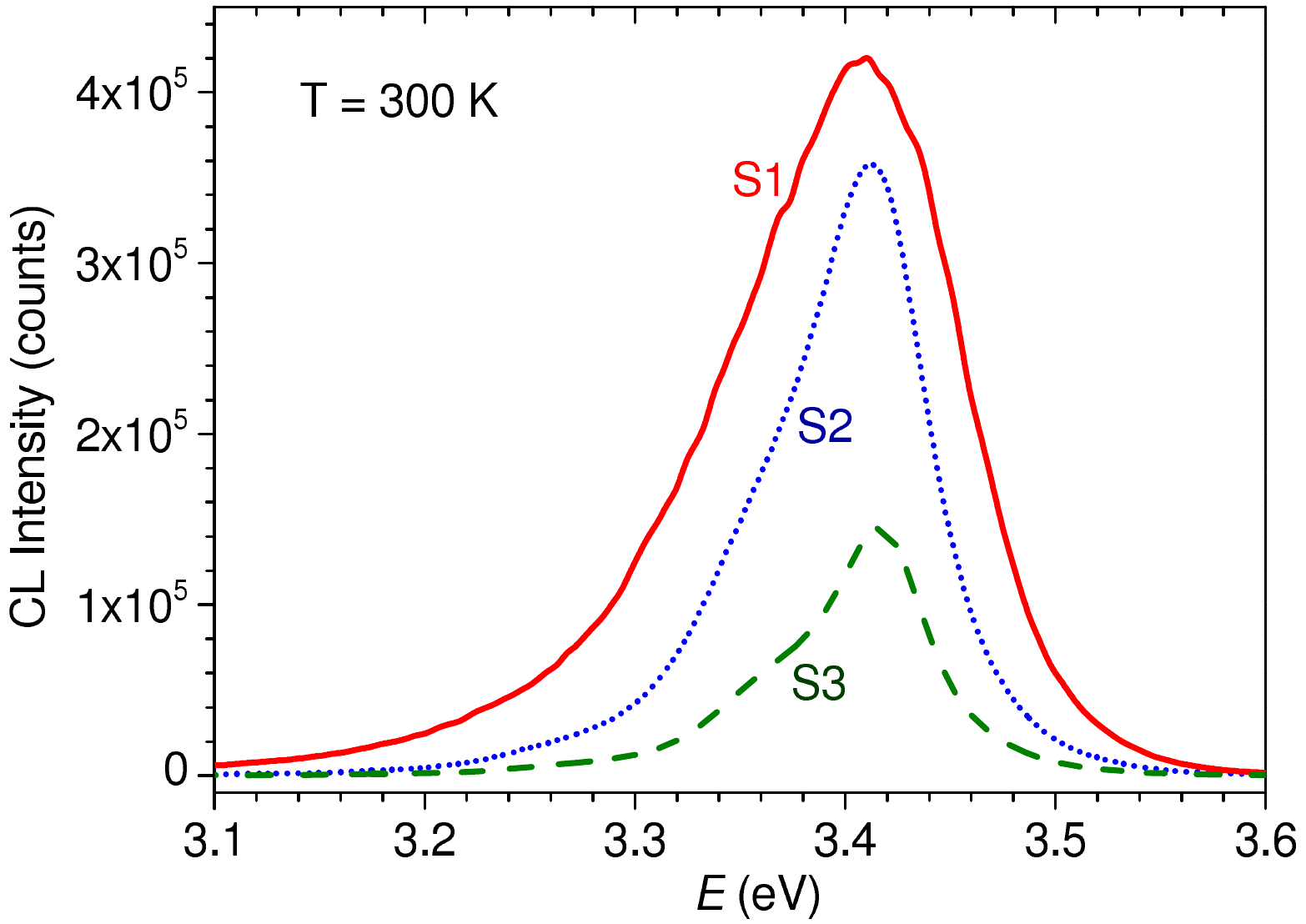}
\caption{Room temperature CL spectra of the upper cross-section regions of the samples S1, S2, and S3 acquired under the same conditions.}
\label{fig9}
\end{figure}

From the comparison of the CL results of S1 and S2 with the ones of S3, we can conclude that the HVPE of GaN on silicon covered by GaN nanostructures in addition to an AlN buffer layer can prevent the incorporation of silicon and results in a higher luminescence efficiency of the HVPE GaN as compared with the growth directly on top of an AlN buffer layer. Moreover, the template of S2 is more suitable for an undisturbed growth along the $c$-direction, which in turn prevents an undesired incorporation of impurity atoms at the growth front.
 
\subsection{Several 100~$\mu$m thick GaN (S4)}
The  conclusion drawn above encouraged us to try a deposition of several 100~$\mu$m GaN using the same template and growth parameters as for S2, but adding a 6 hours long high-$T$ HVPE step at 1063~$^{\circ}$C. As a result, we obtained sample S4 showing a mean thickness of the deposited GaN of 600~$\mu$m. Figure~\ref{fig10}(a) shows an SEM image of the cleaved edge of S4. After the HVPE growth and cooling down process, the silicon substrate was self-separated from the HVPE layer. However, the surface of S4 is apparently rough, and the lower region of the HVPE layer shows a dark contrast in the SEM image. As checked by EDX, this dark part of the HVPE layer is not any more GaN, but a mixture of Ga, N, and silicon. Moreover, the CL maps of figures~\ref{fig10}(b) and 10(c) obtained from an ion-polished region of the cross-section of S4 indicate that the upper, several 100~$\mu$m thick GaN does not exhibit a single, but a polycrystalline structure consisting of several 10 to 100~$\mu$m large crystals. The bright stripes visible in the CL map of figure~\ref{fig10}(c), which are complementary to the luminescence in figure~\ref{fig10}(b), correspond to luminescence of excitons bound to I$_1$-type basal plane SFs and are therefore oriented perpendicular to the $c$-direction. Consequently, the $c$-axies of the grains, which can be recognized in the CL maps, are tilted to each other by several degrees.

\setcounter{figure}{9}
\begin{figure}[h]
\includegraphics*[width=10cm]{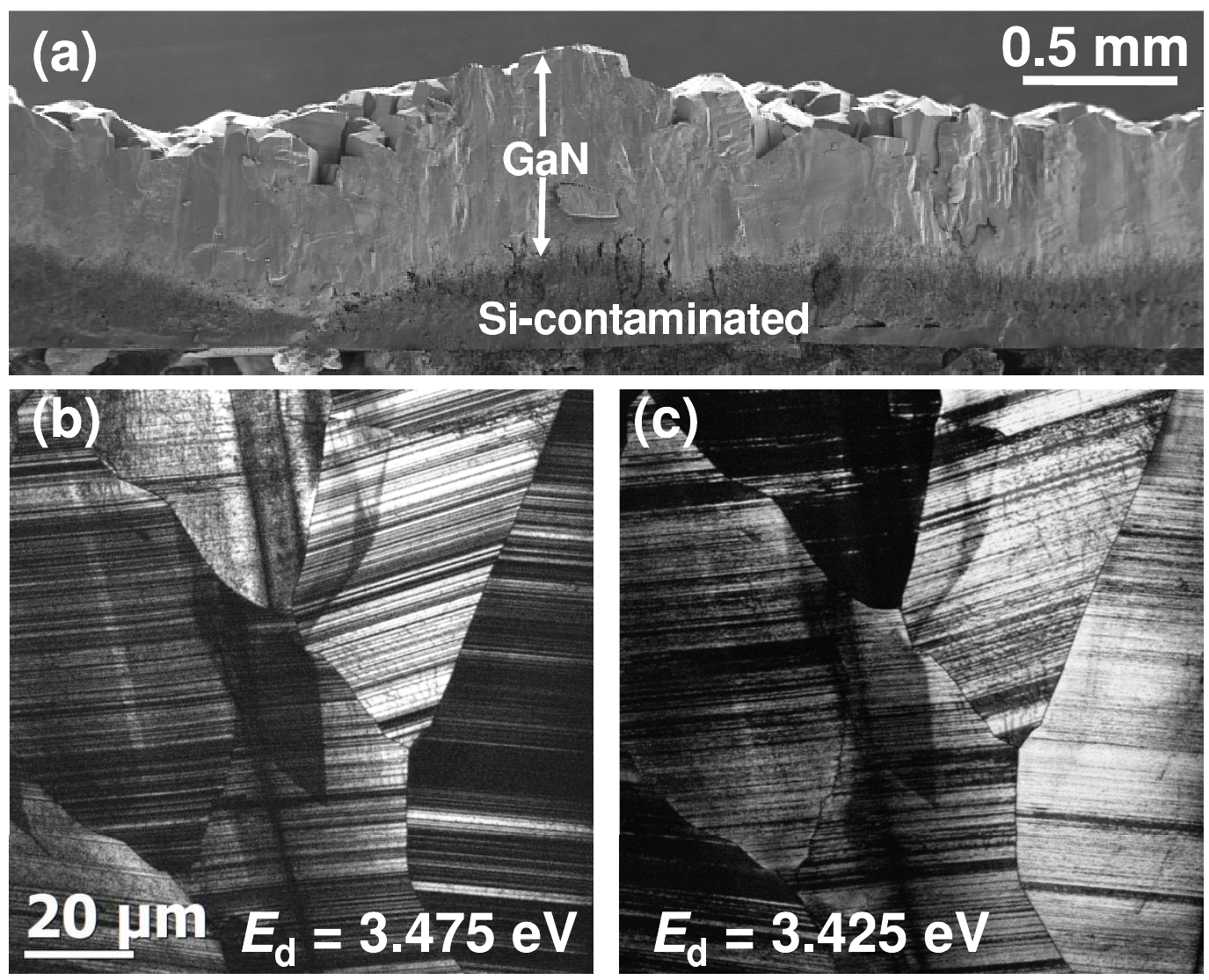}
\caption{(a) SEM image of the cleaved cross-section of sample S4. The silicon substrate was self-separated after the HVPE growth. The regions at the bottom showing a dark contrast are contaminated by silicon. (b) and (c): CL images of an ion beam-polished upper cross-section region of S4, which is free of silicon taken at detection energies of 3.475 and 3.425~eV, respectively.}
\label{fig10}
\end{figure}

The inset of figure~\ref{fig11} shows a large polished piece of S4, whereas the main graph depicts the room temperature CL spectra of this polished piece and the one of the mentioned FS-GaN sample. Both the intensity and the linewidth of the near-band edge CL of S4 are similar to the ones of the FS-GaN sample, where the FWHM amounts to 75~meV. The CL intensity at the spectral range of the "yellow luminescence" (centered at about 2.3~eV) is more than 3 orders of magnitude lower than the near-band edge CL, and therefore, as low as for the FS-GaN. The only significant difference between the two spectra is the appearance of a weak, broad band in the spectrum of S4 centered at about 2.9~eV. The origin of this luminescence band is most probably associated with the presence of a high density of SFs and possibly cubic segments as visible in the low-temperature CL map of the cross-section of S4 in figure~\ref{fig10}(c).

\setcounter{figure}{10}
\begin{figure}[h]
\includegraphics*[width=10cm]{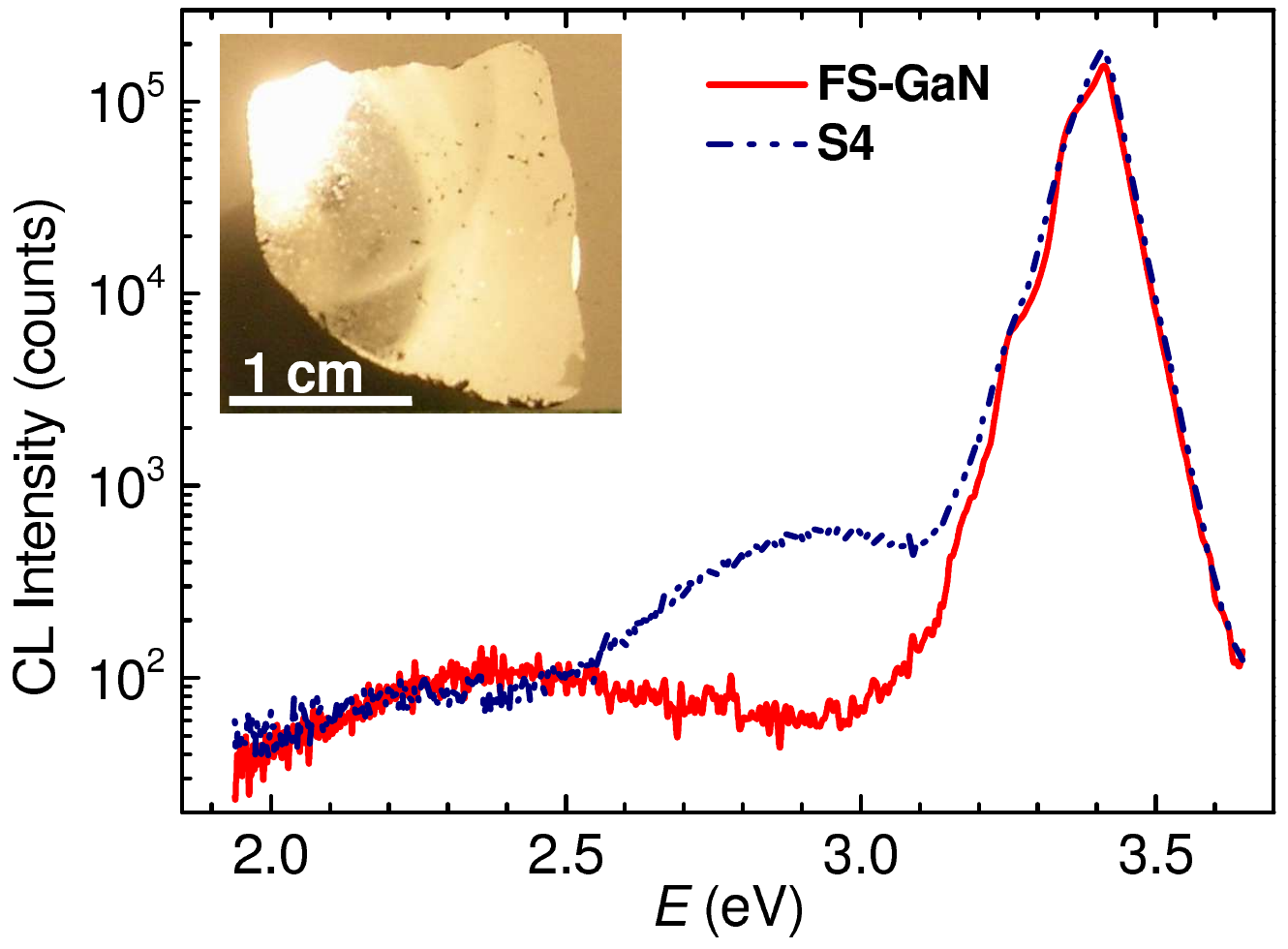}
\caption{Room temperature CL spectra acquired at the surface of a polished piece of S4 as shown in the inset in comparison with a CL spectrum of a free-standing GaN sample.The FWHM of the spectra amounts to 75~meV.}
\label{fig11}
\end{figure}

\section*{Conclusion}
Using a two-step HVPE process, we have grown several ten~$\mu$m thick GaN layers on Si(111) substrates covered by an AlN buffer as protection layer, both without and with additional GaN nanostructures as nucleation sites. A 15~nm thick AlN buffer layer in combination with a low-$T$ HVPE step at 800~$^{\circ}$C can successfully suppress a silicon-melt-back etching. The GaN layer deposited directly on the AlN buffer by HVPE without MBE-grown GaN nanostructures shows a significant broadening and blue-shift of the luminescence spectrum, when excited close to the AlN/GaN interface, which indicates a strong doping of the interface region due to silicon incorporation via fast diffusion paths provided by grain boundaries of the AlN buffer.
When additional GaN nanostructures are deposited by MBE on the AlN buffer, the nucleation and growth during the subsequent low-$T$ HVPE step results in a columnar growth of the GaN, where the column diameter is controlled by the morphology of the nanostructure. Due to the relatively low growth temperature as well as to the lateral growth after nucleation, the GaN layer region of the low-$T$ HVPE step contains a high density of stacking faults, which are no longer present within the region of the high-$T$ HVPE growth. When the diameters of the columns formed during the low-$T$ HVPE step amount to a few 100~nm, the growth dynamics of the subsequent high-$T$ HVPE step is characterized by the formation of V-pits at a very early stage, which in turn is associated with an incorporation of impurity atoms. For diameters of the columns formed during the low-$T$ HVPE step on the order of a few $\mu$m, the growth dynamics of the subsequent high-$T$ HVPE step is characterized by straight columnar growth along the $c$-axis throughout the whole GaN layer. Room temperature luminescence spectra of such GaN layers are very similar to spectra obtained from FS-GaN with respect to both intensity and FWHM.

A several hours long high-$T$ HVPE step resulted in several hundred~$\mu$m thick GaN layers, which show, however, a silicon contamination at their bottom part, a high density of SFs throughout the whole layer, and a polycrystalline structure. Our future research will be focused on the origin of the silicon incorporation during a long-time, high-$T$ HVPE process and the necessary template structure as well as growth parameters to prevent the reaction of silicon with the HVPE GaN.
\

\ack
We thank Oliver Brandt for fruitful discussions and Raffaella Calarco as well as Holger T. Grahn for a careful reading of the manuscript. Moreover, we thank O. Romanyuk for the investigation of the polarity of the HVPE GaN by photoelectron diffraction and we gratefully acknowledge financial support of this work by the Sino-German Center for Science Promotion through the project GZ 736.

\end{document}